# From cold resistor to secure key exchanger

Jiaao Song [1] and Laszlo B. Kish

*Department of Electrical and Computer Engineering, Texas A&M University, TAMUS 3128, College Station, TX 77843-3128*

**Abstract.** Utilizing a formerly published cold resistor circuitry, a secure key exchange system is conceived and explored. A circuit realization of the system is constructed and simulated. Similarly to the Pao-Lo key exchanger, this system is secure in the steady-state limit but crackable in the transient situations.

**Keywords:** Cold resistor circuitry; Pao-Lo key exchanger; Transients; Classical information.

## 1. Introduction

*1.1. On cold resistors*

A *cold resistor* (CR) is an active device, see its black box model in Figure 1. At proper conditions, it behaves as a resistor with $R_{\text{eff}}$ effective resistance and $T_{\text{eff}}$ effective noise temperature that is lower than the ambient temperature $T$. Here $T_{\text{eff}}$ is defined by the Johnson-Nyquist noise formula [1]:

$$T_{\text{eff}} = \frac{S_u(f)}{4kR_{\text{eff}}} , \qquad (1)$$

where $S_u(f)$ is the power density spectrum of the thermal noise voltage of the CR, $k$ is the Boltzmann constant.

---

[1] Corresponding Author

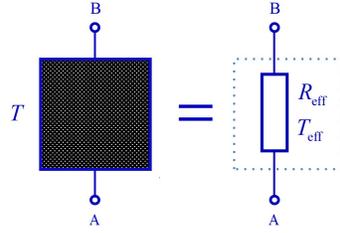

**Figure 1**. The CR circuit as a black box. $T$ is the ambient temperature. $R_{eff}$ is the effective (dynamical) resistance and $T_{eff}$ is the effective noise temperature of the CR which is less than the ambient temperature.

This concept was first introduced many years ago [2]. Subsequently, different cold resistor designs were published [3-8]. Our approach [6-8] is based on the fact that the output resistance of an amplifier and an inserted voltage at its output are reduced by different factors during negative feedback.

In the left section of Figure 2, a serial resistor with resistance $R_0$ is placed at the output of the amplifier of gain $A_1$. Driven by this expanded output (point B), the second amplifier of gain $A_2$ closes the negative feedback loop to the input of the first amplifier. We assume that the amplification of one of the amplifiers is negative, while that of the other one is positive, thus the loop-amplification $A_1 A_2$ is negative:

$$A_1 A_2 < 0 \ . \tag{2}$$

Then the absolute value of the loop-amplification is

$$A_L = |A_1 A_2| = - A_1 A_2 \ . \tag{3}$$

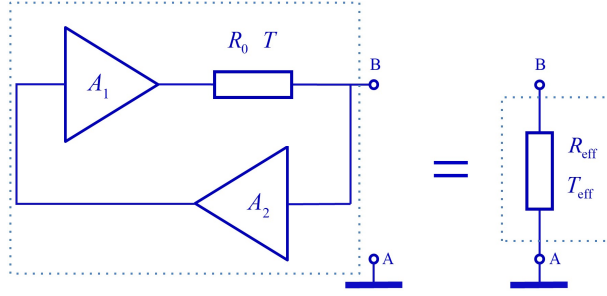

**Figure 2**. Cold resistor circuit with cold output resistance. $A_1$ and $A_2$ are the amplifications of the first and second amplifier, respectively. $R_0$ is the resistance of the inserted serial resistor at the output of the first amplifier. In the box on the right, $R_{\text{eff}}$ is the effective output resistance of the CR on the left. $T_{\text{eff}}$ is the effective noise temperature of $R_{\text{eff}}$ which is lower than the ambient temperature $T$.

The effective resistance $R_{\text{eff}}$ of the CR is the dynamical resistance of the equivalent circuit between point B and ground (point A), which is

$$R_{\text{eff}} = \frac{R_0}{1 + A_L}, \qquad (4)$$

where $R_0$ is the resistance of the inserted serial resistor. From Equation 1, the effective noise temperature is

$$T_{\text{eff}} = \frac{T}{1 + A_L}. \qquad (5)$$

It is known that cold resistors were applied in different research areas, such as gravitational wave detection, photo-amplifiers and quartz oscillators [3-5]. In the subsequent sections, we show a modified system by expanding the cold resistor scheme. At certain situations, the new system shows secure performance.

*1.2. On secure communications*

Communication security is the prevention of unauthorized access to data traffic [9, 10].

The common way to secure a communication is through encryption, that is by generating and transferring a ciphertext obtained by encrypting the plaintext by a cipher [11]. Only authorized parties have the secure key that they can use to decipher a ciphertext and access the information.

For symmetric-key encryption algorithm, which is our focus topic, the same secure keys are applied by both communicating parties [12], see Figure 3.

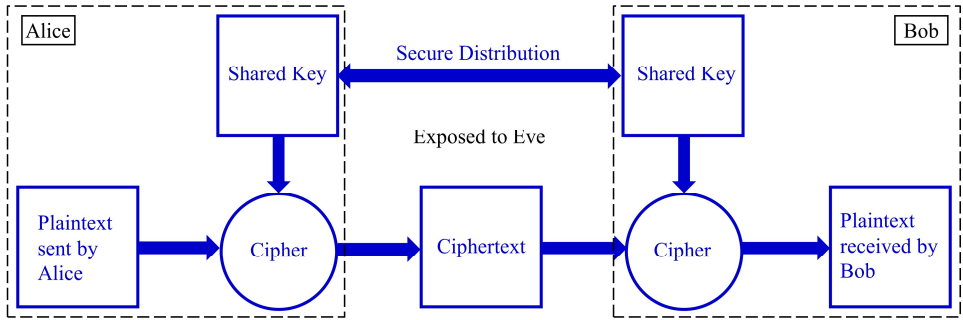

**Figure 3**. Outline of symmetric-key encryption. Alice encrypts the plaintext with a cipher and the shared secure key to generate the ciphertext. Bob uses the same shared key to decrypt the ciphertext with his cipher to get the plaintext. The ciphertext in the communication channel is exposed to Eve, however she does not have the key to decipher it. To generate and share the secure key, a secure key exchange protocol is required.

The challenge for the symmetric encryption algorithm is that Alice and Bob need to generate and share the secure key that itself requires a proper secure communication method: the secure key exchange protocol.

*1.2.1 On the security classes of key exchange protocols*

The security of the key exchange protocol is classified as either conditional security or unconditional security.

Conditionally secure key exchange is typically algorithm-based and uses a mathematical problem. The condition for security is that the cracking of the exchanged key by utilizing Eve's recorded key exchange data is a hard (exponentially complex) computational problem. The system is secure only because Eve has limited (that is polynomial) computational resources and/or limited time. With the development of algorithms and computers, or if Eve has sufficiently long time, she may be able to crack these types of key exchange systems. Conditional security is not future-proof.

Unconditionally (that is, information-theoretically) secure key exchange systems are future proof because Eve's available data do not have enough information to extract the key. Therefore, increased computation power offers no help for Eve. Such key exchangers are typically hardware based. One example is the quantum key distribution (QKD) [13]. The law of physics that provides the security is formulated by the quantum no-cloning theorem that is the impossibility to clone a photon without errors. There have been much debates [14-21] about the foundation of quantum-based security and the difficulty of approaching perfect security with practical quantum hardware.

A classical physical alternative for unconditionally secure hardware is the Kirchhoff-Law-Johnson-Noise (KLJN) key distribution method [22-35]. It is based on the statistical physical features of the thermal noise of resistors and the second law of thermodynamics. Comparing with the QKD method, the KLJN key distribution method has advantages: robustness, low price, the possibility for integration on a chip, etc. [36-42].

*1.3. On the KLJN key distribution system*

The core KLJN key distribution scheme contains two identical resistor pairs (with resistance values $R_L$ and $R_H$, $R_L < R_H$) at the communicating parties Alice and Bob, and the communication channel, which is a wire, see Figure 4. $U_{L,A}(t)$ and $U_{H,A}(t)$ are the Johnson noise of resistors $R_L$ and $R_H$ at Alice's side, respectively. Similarly, $U_{L,B}(t)$ and $U_{H,B}(t)$ are the Johnson noise of resistors $R_L$ and $R_H$ at Bob's side, respectively. $U_w(t)$ and $I_w(t)$ are the wire voltage and current, respectively.

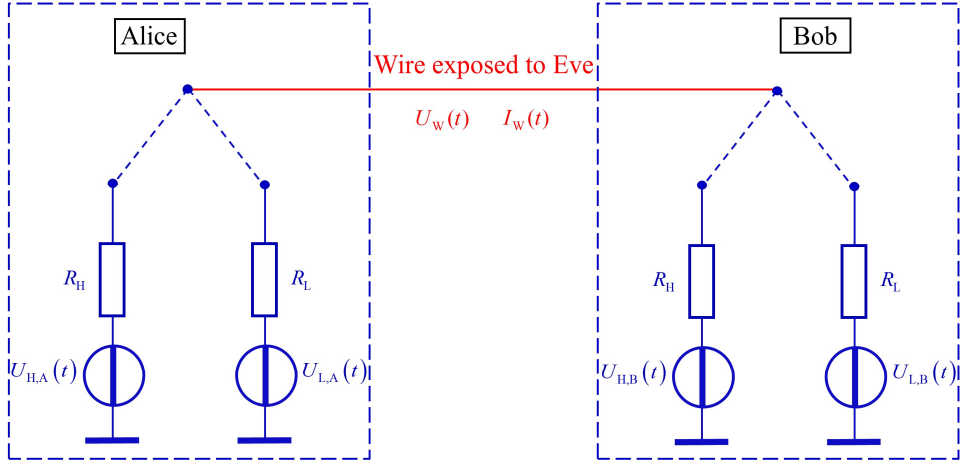

**Figure 4**. The core KLJN key distribution scheme. The two identical pairs of resistors (with resistance values $R_L$ and $R_H$, $R_L < R_H$) located at the communicating parties Alice and Bob, who are connected via a wire. $U_{L,A}(t)$ and $U_{H,A}(t)$ are the Johnson noise of resistors $R_L$ and $R_H$ at Alice's side, respectively. Similarly, $U_{L,B}(t)$ and $U_{H,B}(t)$ are the Johnson noise of resistors $R_L$ and $R_H$ at Bob's side, respectively. $U_w(t)$ and $I_w(t)$ are the wire voltage and current, respectively.

At the beginning of the bit exchange period (BEP), Alice and Bob randomly and independently choose one of their resistors and connect them to the wire. During the BEP, they measure the mean-square current and voltage on the wire.

According to the Johnson formula, the mean-square voltage on the wire is

$$\langle U^2_w(t) \rangle = 4kT_{\text{eff}} R_p \Delta f_B , \qquad (6)$$

where $T_{\text{eff}}$ is the effective noise temperature of the system, $\Delta f_B$ is the bandwidth of the noise generators, $R_p$ is the resistance value of the parallel combination of Alice's and Bob's chosen resistors which is

$$R_p = \frac{R_A R_B}{R_A + R_B}, \qquad (7)$$

where $R_A$ and $R_B$ are the resistance values: $R_L$ or $R_H$ of Alice's and Bob's chosen resistors, respectively.

Based on Equations 6 and 7, there are four different connected resistor combinations: LL, LH, HL and HH are possible, where the first letter refers to Alice's connected resistor and the second letter refers to Bob's. Figure 5 illustrates that resulting mean-square voltages. The LL and HH situations are insecure situations because their mean-square voltages are distinct. Eve can easily identify the connected resistor combination by evaluating the mean-square voltage. Therefore, these insecure exchange bits are disregarded. For LH and HL situations, the mean-square voltages are identical. Eve knows that the situation is either HL or LH but she is uncertain which one. On the other hand, Alice and Bob can distinguish between these combinations because they know their own connected resistors. Therefore, LH and HL represent the two possible values of a shared secure bit.

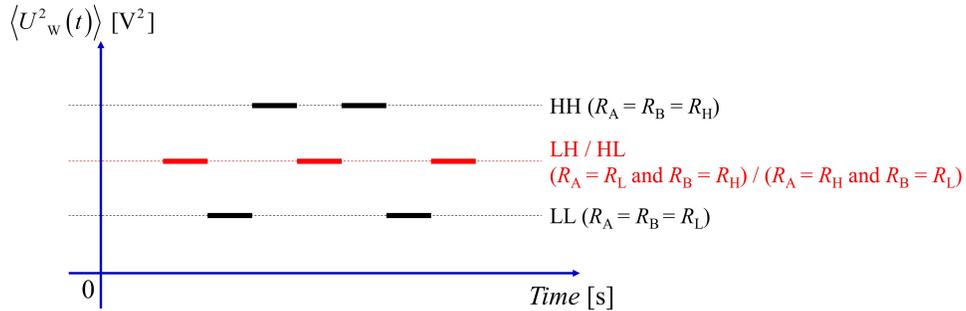

**Figure 5.** Illustration of the mean-square voltage $\langle U^2_w(t) \rangle$ versus clock time during operation. There are four possible connected resistor combinations in the system. The LL and HH situations are insecure because their mean-square voltages are distinct. The LH and HL situations are secure because their mean-square voltages are identical.

In conclusion, the key exchange protocol of the secure bits in the KLJN system is:

- Alice and Bob randomly pick the resistance value $R_L$ or $R_H$ at the beginning of the BEP and contacts these resistors to the wire.
- By executing voltage measurements, and using Figure 5, they guess the resistance value on the other side. That tells them the bit situation at the two ends (LL, LH, HL or HH).
- In the case of LL or HH bit situations, they discard the data.

- They use a formerly agreed truth table about the bit interpretation of the LH/HL resistor combinations to extract the secure key bit. For example, they can publicly agree that LH translates to key bit value 0; and HL translates to key bit value 1.

After this preparation, in Section 2, we show a security related system created from the cold resistor scheme.

## 2. Cold resistor scheme expanded to secure key exchanger

In the left section of Figure 2, by adding the thermal noise voltage source of the resistor, we get the circuit shown in Figure 6.

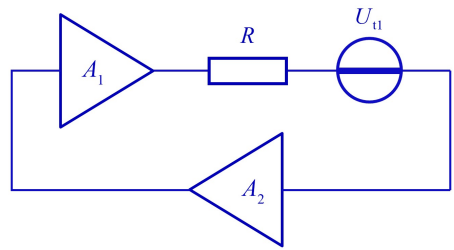

**Figure 6**. Cold resistor circuit (scheme) with the noise generator. $A_1$ and $A_2$ are the amplifications of the first and second amplifiers. $R$ is the serial resistor at the output of the first amplifier. $U_{t1}$ is the thermal noise voltage of the resistor.

We expand and symmetrize the scheme in Figure 6 by adding another resistor $R$ and its thermal noise voltage source $U_{t2}$ between the output of the second amplifier and the input of the first amplifier, see Figure 7.

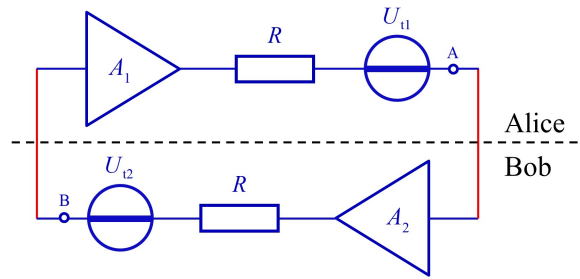

**Figure 7**. The expanded cold resistor scheme. The resistor $R$ and its thermal noise voltage source $U_{t1}$ are in serial at the output of the first amplifier with amplification $A_1$. Point A is the "new output" of the first amplifier that is connected to the input of the second amplifier. Symmetrically, the resistor $R$ and its thermal noise voltage source $U_{t2}$ are in serial at the output of the second amplifier with amplification $A_2$. Point B is the "new output" of the second amplifier that is connected to the input of the first amplifier. The red lines represent the wire-pair connection between Alice and Bob.

In Figure 7, point A is the "new output" of the first amplifier that is connected to the input of the second amplifier. Symmetrically, point B is the "new output" of the second amplifier that is connected to the input of the first amplifier. This symmetrized scheme of the cold resistor scheme inspires us to explore a potential security related feature. Suppose one amplifier is at Alice's side and the other amplifier is at Bob's side. Alice and Bob are able to control their own amplification independently. Below, we explore the security of this scheme.

*2.1. Physical base*

Suppose the delays in the system are negligible. The communicating parties Alice and Bob are connected via two wires, see Figure 8. The mean-square values of the independent thermal noise voltages, $U_{t1}(t)$ and $U_{t2}(t)$ respectively, are equal:

$$\langle U^2_{t1}(t) \rangle = \langle U^2_{t2}(t) \rangle. \tag{8}$$

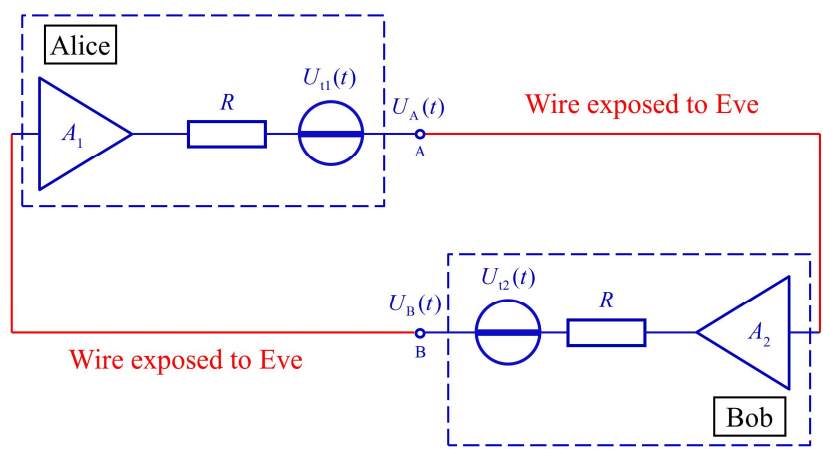

**Figure 8**. Secure system created from the cold resistor scheme. Alice and Bob are connected via two wires that are exposed to Eve. The amplifications $A_1$ and $A_2$, which one fixed during the bit exchange period, are randomly and independently controlled by Alice and Bob. $U_{t1}(t)$ and $U_{t2}(t)$ are the Johnson noises of the resistors at Alice's and Bob's sides respectively and their mean-square values are equal.

In the steady state situation (after the transients have decayed) the following analysis can be made:

$$U_A(t) = A_1 U_B(t) + U_{t1}(t) \tag{9}$$

and

$$U_B(t) = A_2 U_A(t) + U_{t2}(t). \tag{10}$$

From Equations 9 and 10, we get

$$U_A(t) = \frac{U_{t1}(t) + A_1 U_{t2}(t)}{1 - A_1 A_2} \tag{11}$$

and

$$U_B(t) = \frac{U_{t2}(t) + A_2 U_{t1}(t)}{1 - A_1 A_2}. \tag{12}$$

From Equations 11 and 12:

$$\langle U^2_A(t) \rangle = \frac{\langle U^2_{t1}(t) \rangle + A^2_1 \langle U^2_{t2}(t) \rangle + 2 A_1 \langle U_{t1}(t) U_{t2}(t) \rangle}{(1 - A_1 A_2)^2}, \tag{13}$$

$$\langle U^2_B(t) \rangle = \frac{\langle U^2_{t2}(t) \rangle + A^2_2 \langle U^2_{t1}(t) \rangle + 2 A_2 \langle U_{t1}(t) U_{t2}(t) \rangle}{(1 - A_1 A_2)^2}, \tag{14}$$

and

$$\langle U_A(t) U_B(t) \rangle = \frac{A_2 \langle U^2_{t1}(t) \rangle + A_1 \langle U^2_{t2}(t) \rangle + (1 + A_1 A_2) \langle U_{t1}(t) U_{t2}(t) \rangle}{(1 - A_1 A_2)^2}. \tag{15}$$

Because the random noise generator voltages $U_{t1}(t)$ and $U_{t2}(t)$ are independent,

$$\langle U_{t1}(t) U_{t2}(t) \rangle = 0. \tag{16}$$

From Equations 13-16:

$$\langle U^2_A(t) \rangle = \frac{\langle U^2_{t1}(t) \rangle + A^2_1 \langle U^2_{t2}(t) \rangle}{(1 - A_1 A_2)^2}, \tag{17}$$

$$\langle U^2_B(t) \rangle = \frac{\langle U^2_{t2}(t) \rangle + A^2_2 \langle U^2_{t1}(t) \rangle}{(1 - A_1 A_2)^2}, \tag{18}$$

and

$$\langle U_A(t) U_B(t) \rangle = \frac{A_2 \langle U^2_{t1}(t) \rangle + A_1 \langle U^2_{t2}(t) \rangle}{(1 - A_1 A_2)^2}, \tag{19}$$

where $\langle U^2_{t1}(t) \rangle = \langle U^2_{t2}(t) \rangle$.

*2.2. The key exchange protocol: non-secure and secure bit situations*

To use this scheme for secure key exchange, at the beginning of the BEP, Alice and Bob independently and randomly choose their amplification values from $\{A_L; A_H\}$, where $A_L = -A_H$ and $|A_L| = |A_H|$. Suppose $A_L < 0$ and $A_H > 0$. Similarly to the KLJN system, there are four different loop-amplifications (or bit situations): LL, LH, HL and HH, where LL means $A_1 = A_2 = A_L$; LH means $A_1 = A_L$ and $A_2 = A_H$; HL means $A_1 = A_H$ and $A_2 = A_L$; and HH means $A_1 = A_2 = A_H$. Alice and Bob agree about a truth table that (for example) LH and HL mean bit values 0 and 1, respectively.

Eve has access to the wire connections and can measure the cross-correlation $\langle U_A(t)U_B(t) \rangle$. According to Equation 19, for LL, $\langle U_A(t)U_B(t) \rangle < 0$. For HH, it is $\langle U_A(t)U_B(t) \rangle > 0$, see Figure 9. Thus, Eve can identify LL and HH as they are unsecure situations. It's worth mentioning that for these two situations, the loop-amplifications, $\hat{A}_L^2$ and $\hat{A}_H^2$, respectively, are positive, thus the system can become unstable at certain situations. This is another argument why the LL and HH cases are useless and these bits must be discarded.

For LH and HL, according to Equation 19, we know that the cross-correlations of $\langle U_A(t)U_B(t) \rangle$ are identical and equal to zero (Figure 9). By measuring and finding the cross-correlation zero, Eve knows that, at one of the sides, the amplification is $A_L$ and, at the other side, the amplification is $A_H$. However, she has no information about the exact location of these amplifications. Thus, Eve cannot learn the corresponding secure bit value. On the other hand, Alice and Bob can determine the amplification at the other side with the knowledge of the amplification values at their own sides. Similarly to the KLJN scheme, see Section 1.3, using the above specified truth table for the LH and HL situations, a secure bit is shared between Alice and Bob.

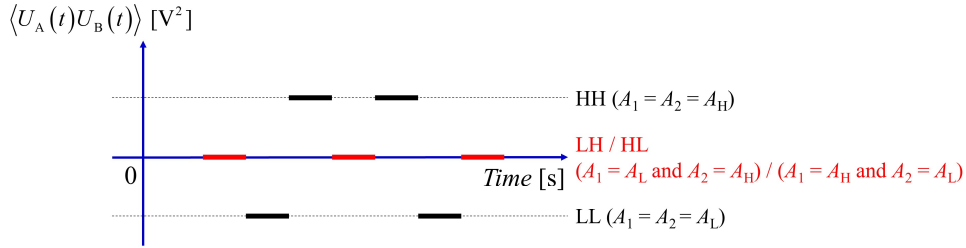

**Figure 9**. Example of the cross-correlation $\langle U_A(t)U_B(t) \rangle$ versus clock time during operation, where $A_L < 0$, $A_H > 0$ and $|A_L| = |A_H|$. There are four different situations. Similarly to the KLJN key exchange system, the LL and HH situations are insecure because the mean voltages are distinct. The LH and HL situations are secure because the cross-correlations are identical (zero).

In conclusion, Alice and Bob use the following protocol to extract the secure bits:
- They independently and randomly choose the amplification values $A_L$ or $A_H$ at the beginning of the BEP, where $A_L < 0$, $A_H > 0$ and $|A_L| = |A_H|$.
- By executing the cross-correlation measurements, and using Figure 9, they determine the amplification value at the other side. That tells them the bit situation at the two ends (LL, LH, HL or HH).
- In the case of LL or HH bit situations, they discard the data.
- In the case of LH or HL bit situations, similarly to the KLJN key exchanger, they use a formerly agreed truth table about the bit value interpretation of the amplification situation to extract the secure key bit.

On the average, every second clock period yield a shared secure bit (LH or HL). The other half of results are discarded.

### 2.3. On external noise generators

In Figure 8, the random noise voltage generators represent the Johnson (thermal) noise voltages of the resistors in the system, with noise spectrum

$$S_{u,\text{th}}(f) = 4kTR, \qquad (20)$$

where $T$ is the ambient temperature.

However, it is more practical to substitute these random noise voltage generators by external noise voltage generators which follow the same scaling law between the resistance and the mean-square voltage noise as the thermal noise. Then much higher noise temperatures $T_{\text{eff}}$ (and noise amplitude) can be used. The noise voltage spectrum is

$$S_{u,\text{ex}}(f) = 4kT_{\text{eff}}R, \qquad (21)$$

where the $T_{\text{eff}}$ can be chosen sufficiently large for practical application to eliminate the problems with parasitic background noises.

### 2.4. The Pao-Lo key exchanger

Before showing our test results, due to its relevance, we briefly mention a former key agreement protocol using band-limited random signals and feedback proposed by Pao-Lo Liu [43, 44]. We call it Pao-Lo key exchanger. It is the software-based emulation of the KLJN secure key exchanger by emulating stochastic light sources and mirrors, which can have positive or negative reflectance value. Figure 10 shows the diagram of Pao-Lo key exchange system.

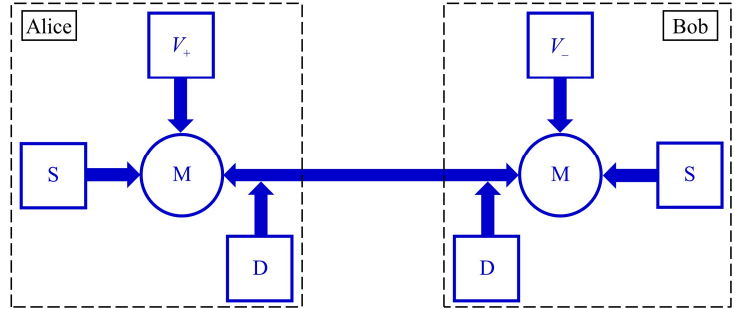

**Figure 10**. Diagram of Pao-Lo key exchange system [43, 44]. **S** represents the stochastic light source. **M** represents the controlled mirror. **D** represents the detector. $V_+$ and $V_-$ are the band-limited signals transmitted by Alice and Bob respectively.

We note that our scheme introduced in Section 2 is a kind of the circuit realization of the Pao-Lo key exchanger. The reflectivity corresponds to the amplification. The two wires represent the two propagation directions.

## 3. Test results

Below we show some of the results of the simulation results carried out by the *LTspice XVII* industrial circuit tester/simulator system.

### *3.1. The simulated circuitry*

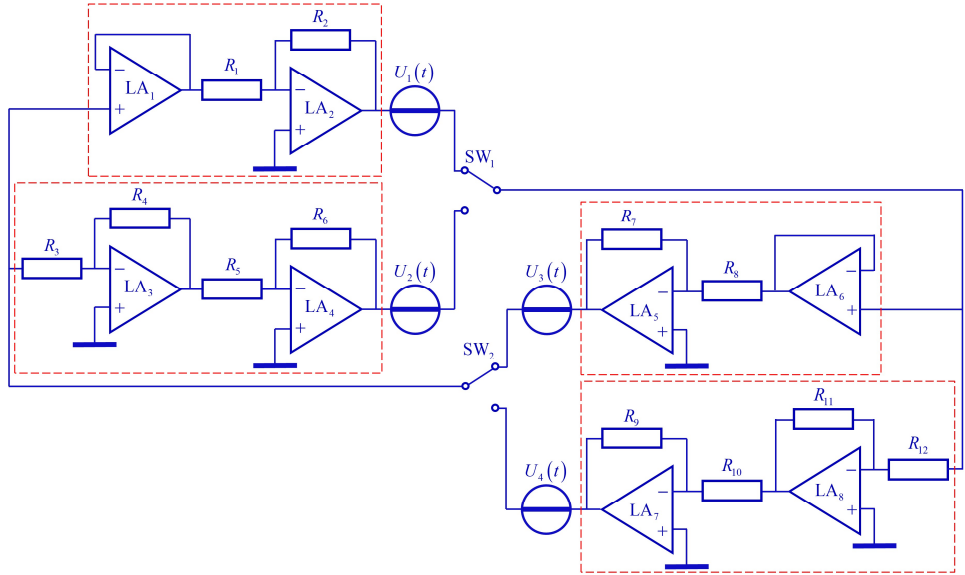

**Figure 11**. The circuitry of the key exchange system. $LA_1$, $LA_2$, $LA_3$, $LA_4$, $LA_5$, $LA_6$, $LA_7$ and $LA_8$ are ADA4627 linear operational amplifiers. Resistances $R_1$, $R_2$, $R_3$, $R_4$, $R_5$, $R_6$, $R_7$, $R_8$, $R_9$, $R_{10}$, $R_{11}$ and $R_{12}$ are properly chosen to control the amplifications. $U_1(t)$, $U_2(t)$, $U_3(t)$ and $U_4(t)$ are random noise voltage generators. $SW_1$ and $SW_2$ are ADG1219 SPDT controlled analog switches.

For the circuitry (Figure 11), we chose $R_1 = R_2 = R_5 = R_6 = R_7 = R_8 = R_9 = R_{10} = 100$ Ohms. $R_3 = R_4 = R_{11} = R_{12} = 500$ Ohms. $LA_1$, $LA_2$, $LA_3$, $LA_4$, $LA_5$, $LA_6$, $LA_7$ and $LA_8$ are ADA4627 operational amplifiers. $SW_1$ and $SW_2$ are ADG1219 SPDT analog switches independently controlled by square wave voltage sources with 50% duty cycle providing the clock signals for the bit exchange periods of 1.25 ms. Their waveform has 2.5 ms period and 3 V amplitude. The random noise voltage sources $U_1(t)$, $U_2(t)$, $U_3(t)$ and $U_4(t)$ are standard Gaussian random signals with 250 kHz sampling rate. These noise signals are filtered by 5 kHz low-pass filters before injecting them into the circuitry.

## 3.2. The secure (steady-state) situation

The steady-state results are shown in Figure 12. The 1.25 ms BEP is also the averaging time of the cross-correlation in Equation 19. Due to this finite averaging time, there are statistical inaccuracies resulting in the scattering of data. The LL and HH situations can be distinguished and are insecure because their data ranges are distinct. On the other hand, for LH and HL situations, the data are randomly scattered around zero in the range of -0.03 to 0.03 $V^2$. The randomness of this scattering around zero guarantees the security.

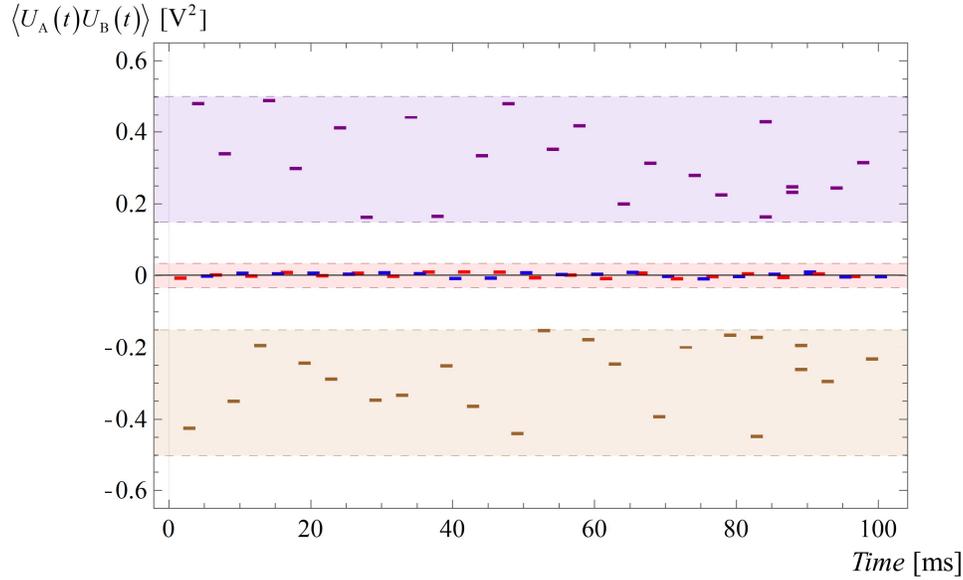

**Figure 12**. The cross-correlation $\langle U_A(t)U_B(t) \rangle$ versus clock time by simulation at BEP=1.25 ms and noise bandwidth 5 kHz. The purple lines represent the cross-correlations in the HH situation. The brown lines represent the cross-correlations in the LL situation. The red and blue lines represent the cross-correlations in situations LH and HL, respectively. They are randomly scattered around zero in the range of -0.03 to 0.03 $V^2$. The randomness of this scattering around zero guarantees the security.

## 3.3. Transients: security vulnerability

Similarly to the Pao-Lo key exchanger [43, 44], the security vulnerability of the system is due to the transients between different bit situations, that is, between different amplification combinations. They are due to the abrupt switching changes and the nonlinear dynamical transient performance of the circuitry.

For the transient states, in total, there are 12 different situations: LL→HL, LL→LH, LL→HH, LH→HH, LH→LL, LH→HL, HH→LH, HH→HL, HH→LL, HL→LL, HL→HH and HL→LH. We know LL and HH are insecure bit situations thus they are discarded. Therefore, the transitions LL→HH and HH→LL are also insecure and discarded.

Our extensive simulations indicate that each of these transitions leak significant information to Eve. Due to the limitation of space, here we show only a few characteristic transient examples as demonstration to demonstrate the security vulnerability.

There are large variations of transients versus the initial voltages before the transient, see Figures 13, where (a) to (f) are different realizations of the voltages in the wires at the HL→LL transition. $V_{AB}$ is the transient voltage in the wire from Alice's output to Bob's input and $V_{BA}$ is the transient voltage in the other wire. These results indicate that the shape of the transients depends not only the logic transition but also on the initial voltage levels before the transient.

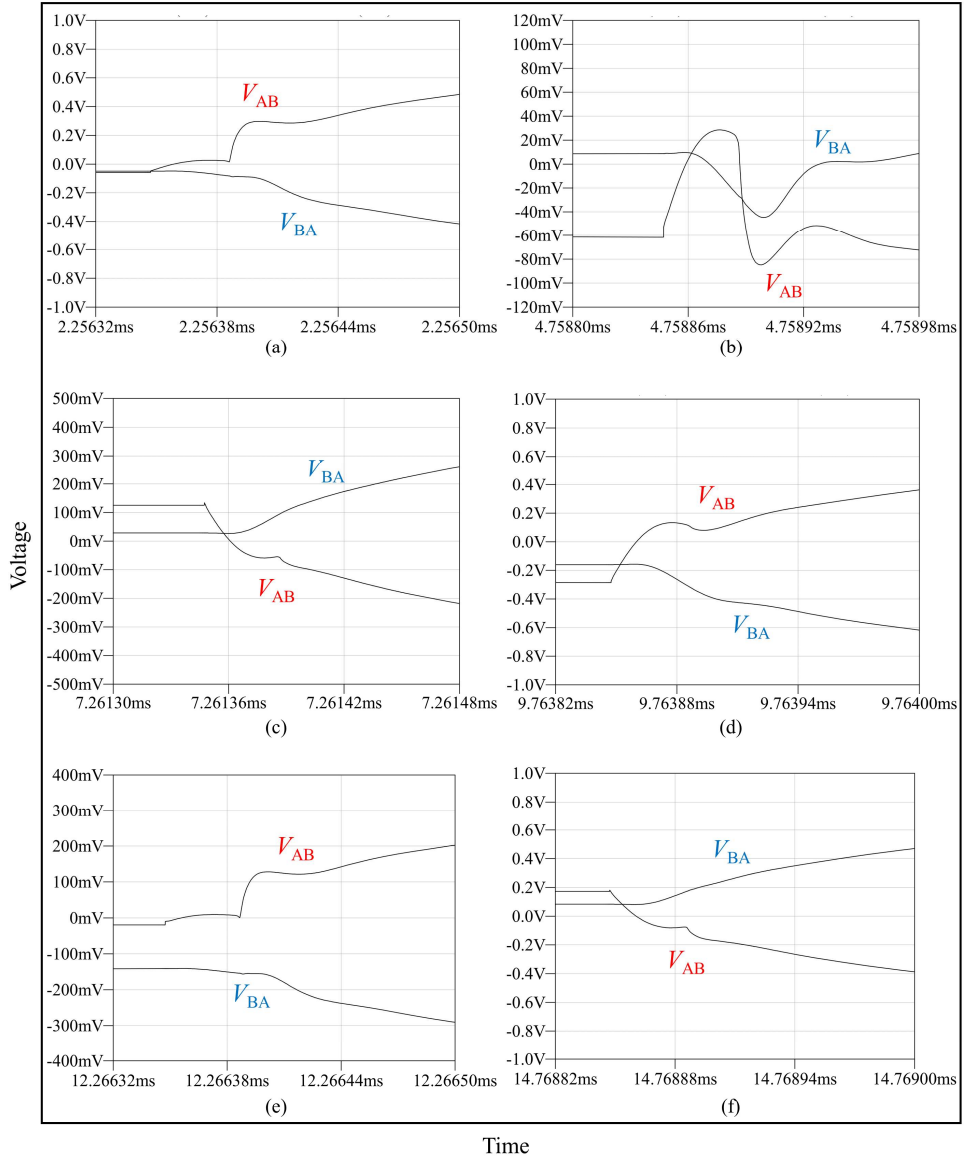

**Figure 13**. (a) to (f) are different realizations of the voltages in the wires at the HL→LL transition. $V_{AB}$ is the transient voltage in the wire from Alice's output to Bob's input and $V_{BA}$ is the transient voltage in the other wire. The shape of the transients depends not only the logic transition but also on the initial voltage levels before the transient.

It is important to realize that the transients are very fast compared to the characteristic time constant (autocorrelation time) of the voltage level of the noise, which is approximately fixed throughout the transient, similarly to the Pao-Lo key exchanger. Thus, the transients are virtually deterministic. Due the exact symmetry of the circuitry, the same transient exists on the alternative wires when the logic transition is replaced by the opposite one, provided that the initial voltage levels are also swapped between the two wires.

Due to the Kerckhoffs's principle of security [22], Eve knows all the fine details of the system of Alice and Bob, including their protocol. Thus, Eve can build the whole circuitry and do extensive simulations of the transients at a large set of various initial conditions. She can create a massive transient database and a relevant truth table to identify the logic transition corresponding to the observed transients in the two wires at a large number of initial conditions.

In Figure 14, an example is shown with two nearly symmetric cases of pre-transient voltage levels. Utilizing the symmetry of the circuitry, Eve can distinguish the HL and LH situations at similar initial voltages. If a sufficiently large database is available with transients at various initial voltages, Eve can crack the system due to the deterministic nature of these transients which are determined by the initial conditions. The symmetry of the system reduces the size of the necessary database.

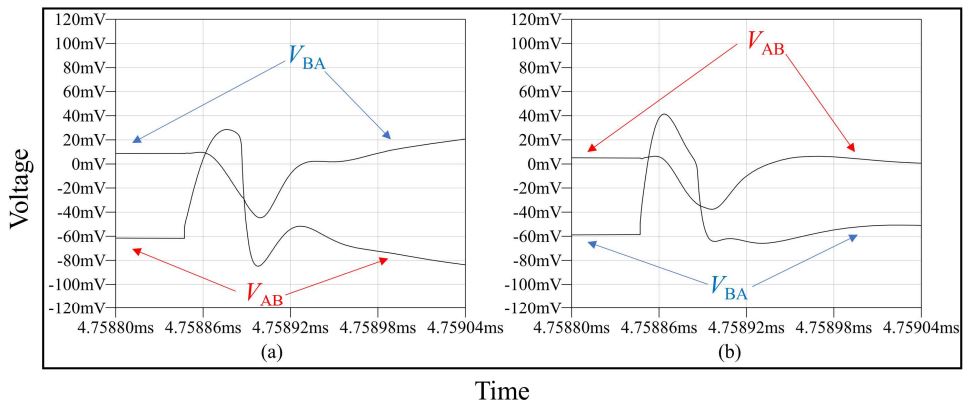

**Figure 14.** Example showing the symmetry and virtual determinism of the transients and their utilization for attack when the approximate pre-transient voltage levels are nearly identical. (a) Simulation of the voltages in the wires at the HL→LL transition. (b) Simulation of the voltages at the LL→HL transition. Eve can create a truth table by collecting a large number of transients with various initial conditions. The initial conditions determine the transients. The symmetry of the system reduces the size of the necessary database.

In conclusion, a transient defense protocol is required to fix the security not only in the steady states but also during the transitions [22, 25]. To make the system secure against such a transient attack, a proper transient protocol should introduce randomness into these transients so that Eve's uncertainty about the meaning of transitions approaches 1 bit information entropy. It is an open question if proper softening of the switching events, as it is proposed in the KLJN key exchanger [29] by ramping up the voltages is enough, or additional precautions are required [22, 25].

Finding such a protocol is the subject of current research and it is out of the scope of this paper.

**Conclusions**

We introduced and explored a cold resistor scheme that is expanded to become a secure key exchanger. A circuit model of the system is constructed and simulated. Similarly to the Pao-Lo key exchanger [43, 44], this system is secure in the steady-state limit but crackable in the transient situations.

It is an open question that what kind of expansions and transient protocol can randomize the transients to maximize Eve's uncertainty about the actual bit situations.